# An Overview on Data Security in Cloud Computing

Lynda Kacha and Abdelhafid Zitouni

Lire Labs, Abdelhamid Mehri Constantine 2 University,
Ali Mendjli, 25000 Constantine, Algeria
lyndakacha@yahoo.fr,
Abdelhafid.zitouni@univ-constantine2.dz

**Abstract.** Cloud Computing refers to the use of computer resources as a service on-demand via internet. It is mainly based on data and applications outsourcing, traditionally stored on users' computers, to remote servers (datacenters) owned, administered and managed by third parts. This paper is an overview of data security issues in the cloud computing. Its objective is to highlight the principal issues related to data security that raised by cloud environment. To do this, these issues was classified into three categories: 1-data security issues raised by single cloud characteristics compared to traditional infrastructure, 2-data security issues raised by data life cycle in cloud computing (stored, used and transferred data), 3-data security issues associated to data security attributes (confidentiality, integrity and availability). For each category, the common solutions used to secure data in the cloud were emphasized.

## 1 Introduction

Nowadays, sales of digital devices (smartphones, and tablets) have exploded. These connected and increasingly mobile devices allow users to access their data and applications from anywhere and anytime. The traditional IT infrastructure is no longer suitable for ubiquitous access to data. It becomes costly and difficult to manage. The rapid increase in global internet usage requires a new way to manage the size, variety and availability of data, which is CLOUD COMPUTING [27].

Cloud Computing introduces a new way of providing resources to users "as a service accessible via the internet". Unlike traditional approach that is based on hardware ownership where data is stored, Cloud Computing users no longer own the infrastructure that is totally controlled by these service providers. The transfer of infrastructure control to the service providers involves the transfer of responsibility associated with data security [1, 4, 14, 19, 31, 34]. Therefore, data security and privacy concerns raise.

Cloud Computing domain is so wide that it is impossible to deal with all of its aspects. In this paper, we focus on aspects related to Cloud Computing security and more particularly, we are interested in the security of data hosted on Cloud infrastructures. As more and more information from individuals and organizations are placed on the Cloud, the issue regarding data security and user privacy becomes an important concern, especially when data is sensitive. The rest of the paper is organized as follows: Sect. 2 discuss the data security issue according to three dimensions- cloud characteristics, data life cycle and data security attributes. We identify in Sect. 3 common solutions to data security issues in cloud according to data security attributes. Section 4 concludes this paper.

## 2 Data Security Issues Classification

Data security is a common concern to all technologies. However, it becomes a major challenge when applied to an uncontrolled environment like Cloud Computing. It is important to distinguish between the security risks associated with all IT infrastructures and those introduced by the use of Cloud Computing. These risks are generally associated with open, shared and distributed environments. Therefore, when analyzing the risks, it is important to separate existing problems from those raised by Cloud Computing. In this paper, we deal only with issues introduced by the Cloud, and related to data.

Data outsourced to Cloud infrastructure is more vulnerable than that stored on a traditional infrastructure, mainly for three reasons: (1) data is stored on the service provider's infrastructure; (2) data of different users shares the same physical infrastructure; (3) data is accessible via internet.

Although there are many possible classifications for data security issues, we have chosen to classify them according to three dimensions: single Cloud characteristics, Cloud data life cycle and data security attributes, as shown in Fig. 1.

Our objective is to highlight the impacts of these dimensions on data security as well as the common and distinct data security implications associated to these categories.

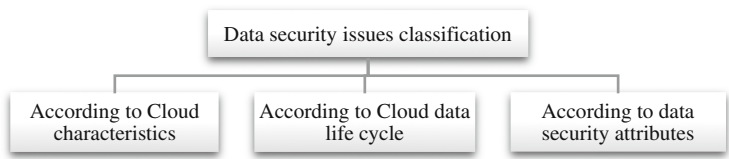

**Fig. 1.** Data security issues classification

### 2.1 Data Issues According to Cloud Characteristics

We are interested, in this section, to data security issues raised by characteristics of Cloud infrastructure compared to traditional proprietary infrastructure. Indeed, Cloud infrastructure is different from traditional infrastructure. This differences offer many benefits but also introduce numerous inconvenient, which may affect security. The main characteristics and theirs direct benefits and inconvenient are the following:

*Leased infrastructure*

Cloud infrastructure no longer belongs to user but to service provider. Instead of purchasing dedicated hardware, users lease its use from a service provider. Principal advantage: cost saving. Principal inconvenient: loss of control.

*Open infrastructure*

Cloud infrastructure is, generally, accessible via internet. Principal advantage: ubiquitous access to services. Principal inconvenient: multiple entry points.

*Shared infrastructure*

Unlike dedicate traditional infrastructure, Cloud infrastructure is shared among service user. Principal advantage: cost saving. Principal inconvenient: Isolation failure risks between users.

*Elastic infrastructure*

Users Cloud can scale up/down the resources according to their need. Therefore, unlike to the traditional infrastructure that depends on peak of demand, Cloud infrastructure scale to current demand. Principal advantage: resource use optimization. Principal inconvenient: resource reallocation risks.

*Virtualized infrastructure*

Virtualization is the basic concept behind the Cloud. We no longer refer to the physical machine but rather to the virtual machine. Principal advantage: infrastructure optimization. Principal inconvenient: Classical problems associated with virtualization.

*Distributed infrastructure*

Cloud infrastructure is distributed geographically around the world. Principal advantage: Increase Computing and storage capacity. Principal inconvenient: Management and maintenance of infrastructure.

It is obvious that these characteristics have an impact on data security. We have excerpted the main data security issues raised by these characteristics, which we have summarized in Table 1.

## 2.2 Data Issues According to Cloud Data Life Cycle

Calculation and storage are the two basic services provided by Cloud Computing [2, 5, 21]. Data storage is distributed over a number of Datacenters around the world. Data calculation is carried out by virtual machines. Users can create different virtual machines, with different capacities and numbers to suit their needs [5]. The transfer of data calculation and storage to a third part involves the transfer of responsibility associated with their security and compliance to this third part [29].

The calculation in the Cloud takes place as follows: the user first submits his data to the datacenter that is stored and managed by storage service. This data is then sent to the virtual machines for parallel processing using the corresponding distributed technology. After the end of processing, users can download and view the results [5]. During this process, all private or confidential data may be disclosed.

**Table 1.** Data security implications according to Cloud characteristics compared to traditional infrastructure

| Principal characteristics | | Advantages | | Data security implications in the cloud |
|---|---|---|---|---|
| Cloud infrastructure | Traditional infrastructure | Cloud infrastructure | Traditional infrastructure | |
| Leased infrastructure | Proprietary infrastructure | Cost reduction- abstraction of hardware and software management constraints, physical security | Better control over infrastructure, more cost-effective when needs are "stable" | Loss of control over data → risks related to data confidentiality, integrity and availability |
| Open infrastructure | Closed infrastructure | High availability | Better security level | Unauthorized access → Risks related to data confidentiality, integrity and availability |
| Shared infrastructure | Dedicate infrastructure | Cost reduction- collaboration between users- optimized management of physical infrastructure | Physical isolation between users | Unauthorized access between Cloud consumers → Risks related to data confidentiality, integrity and availability |
| Elastic infrastructure (scale up/down) | Rigid infrastructure (scale up) | Cost reduction- resources using optimization | Simpler infrastructure management | Risks related to data confidentiality (resource reuse, case of data remanence) |
| Multi-level Virtualization (infrastructure, platform, application) | Virtualization possible, usually on a single level | Cost reduction- optimization and easier maintenance of physical resources- flexibility- simple, fast and dynamic management of virtual resources | Easier infrastructure management- better security | Classical virtualization risks (hypervisor, virtual machines, virtual network, and the problem of sharing physical resources) → Risks on data confidentiality, integrity and availability |
| Distributed infrastructure | Centralized infrastructure | High availability, better fault tolerance | Easier, more controlled and more secure infrastructure management | High risk on data confidentiality and privacy but also on integrity |

Based on this process, we can distinguish three states relating to the data in the Cloud: data-at-rest, i.e. the data stored, data-in-transit i.e. the data transmitted and the data-in-use i.e. data accessed or being processed. Therefore, data security and exploitation in the Cloud must cover these three aspects. At each stage of this data life cycle, different measures can be implemented to ensure data security.

We highlight in this section the data security issues related to each stages of this life cycle. These issues have been extracted from various paper dealing with the subject. We have summarized the main ones in the following:

**Data-at-rest**

Data storage is one of the most commonly used services in the Cloud. It offers the user an "unlimited" space and allows him to access his data ubiquitously at a lower cost. Data-at-rest security refers to securing data on the storage media. It is difficult to achieve for the user due his limited physical control over the data.

There are many risks mentioned in the literature concerning data storage in Cloud Computing. The risks shared by most work have been summarized here in three categories: risks associated with storage media sharing, risks associated with data location and risks related to storage media reliability.

*Risks associated with storage media sharing*

The first risk concerns the sharing of physical infrastructure between different users. Given, multiple users share the same physical storage space, data security risks increase and affect many users. Unauthorized access by a service provider or its customers (other users) is therefore a serious problem, especially, when data is sensitive. Data theft, leakage and alteration are the main risks compounded by storage sharing [13, 14, 24, 28, 30, 31].

*Risks associated with data location*

The second risk concerns physical data location. User data is stored in different locations distributed around the world. In general, Cloud users do not know the exact location of their data; In most cases it does not matter. For example, messages and photos exchanged in Facebook can reside anywhere in the world and Facebook users do not usually worry about that. However, when an enterprise has some sensitive data on the Cloud, the location of the data becomes important. The geographic location of the Cloud provider can have an impact on data security and privacy; indeed, authorities of a country where data resides may access the data under certain circumstances in accordance with the laws of that country [8, 14, 21, 24, 25, 28, 33]. For example, Patriot Act in the United States. Under this Act, FBI and similar organizations have the regulatory authority to access all data stored on any computer in the United States; Even if they belong to other countries [4, 9, 25, 29]. Actually, the majority of Datacenters reside in the United States. Therefore, data protection and privacy are influenced by American laws.

*Risks associated with storage media reliability*

The third risk is the reliability of Cloud storage platform. Since users do not control physical access to data, they must rely on the provider to secure their data. In addition, service provider may subcontract its services to another provider without the client being informed, which can increases the risk

**Data-in-transit**

Data-in-transit security refers to the security of data transmissions in the Cloud. It ensures that the data will not be intercepted, altered or replaced. data-in-transit can be very sensitive like user names and passwords. Data-in-transit may be more at risk than data-at-rest, as they travel from one place to another [16].

**Data-in-use**

Data-in-use refers to any reading or processing (creation, transformation or deletion) of data. When processing take place in the Cloud, the risks of misuse increase, due to the large number of users involved in Cloud [22].

In a traditional environment, the user holds, and manages his data. However, in Cloud Computing a user's data can be generated and handled by a third party. The problem for the owner is to keep control over his data created by another. For personal and private information, the owner must know what personal information is collected, and in some cases, stop collecting and using of this information. Furthermore, Owners of data need to ensure that the use of their data is consistent with the purposes of the collection and that private information is not disclosed to third parties [11, 22].

Data deletion in another problem concerning data processing in the Cloud. Due to the physical characteristics of the storage devices, the deleted data may still exist and can be restored. This problem is called data remanence. It was emphasize by numerous works [14, 18, 22, 28, 33]. This problem arises when data residues are present on the storage media after deletion operation. Data remanence is a serious threat to data privacy, whereby confidential data can be revealed, especially when this data is in an uncontrolled environment such as the Cloud. The question that arises is how can a user ensure the effective deletion of his data when he does not have physical control over the storage media? There is currently no way to prove this, because it is in the first place a problem of trust. This concern increases that there may be multiple copies of the data, potentially held by several entities. There is, therefore, a high risk of data exposure for a user if the physical resources are reused by another user.

### 2.3 Data Issues According to Data Security Attributes

Although the security requirements differ from one data type to another (data-at-rest, data-in-transit, data-at-use). They all share a basic concept that is CIA trio: Confidentiality, Integrity, and Availability, but applied to a distributed, virtualized and dynamic architecture. These three principles are used by all security measures that are intended to protect one or more aspects of this trio. The majority of literature papers [2, 4, 7, 8, 17–19, 21–23, 30, 33] dealing with data security discuss these three points.

**Confidentiality**

Confidentiality refers to data protection from unauthorized access. This problem occurs when sensitive data is outsourced to the Cloud server. In a decentralized Computing context, the issues of confidentiality are much more important since the server hosting the data does not necessarily belong to the user. Confidentiality in Cloud systems is a major barrier to his adoption. Currently, Cloud offers are mainly public and therefore exposed to more attacks, compared to those hosted on private data centers.

Data privacy is another problem often associated with confidentiality. Privacy concerns personal information that must be hidden from unauthorized persons. The user privacy is associated with the collection, use, communication, storage and destruction of personal data. This concern arises when the reasons for use of personal data and the way in which they are used are not clear [11].

**Integrity**

Integrity refers to data protection from unauthorized changes, whether intentional or accidental. These changes include creating, deleting, and writing. Data integrity is one of the critical elements in most information systems. It can be simple to perform in a centralized system, but becomes a complex task in a distributed environment such as Cloud Computing.

An important process is the verification of data integrity. To do this, the backup of the original data is generally compared to the current data in the Cloud. However, this method involves downloading data from the Cloud, which can be complex and costly [2]. Today, there are other techniques that do not involve downloading data to check their integrity.

**Availability**

Data availability means that information must be available when authorized persons need it. Data availability is one of the biggest concerns of service providers. If for some reason a Cloud service is interrupted, many clients will be affected. Service providers contractually Undertake to ensure an availability level of 99.9%. In addition, the duplication of data and physical resources and their distribution on different locations increases the level of availability.

There are many risks that could affect the availability of data in the Cloud such as storage reliability, dependence on internet connection and technical failures. Generally, data availability in Cloud is more reliable than on a traditional infrastructure as large suppliers like Google, Amazon and Microsoft are better equipped to manage these risks than a simple individual or a company.

## 3 Data Security Issues Solutions

Cloud Computing is used in a variety of service models: SaaS, PaaS, IaaS: and deployment models: private, public, hybrid, community. Therefore, the risks are different depending on the level of cloud used; indeed, if the security control on a private Cloud is logically high since mastered, the level of control over a public Cloud is substantially lower. Likewise, whether the user depends on a software, platform or infrastructure, the level of control is different and thus the security management will be different. IaaS provides an infrastructure to host PaaS, which in turn provides a platform for developing and deploying SaaS applications; therefore, there is a security dependency between these layers.

Moreover, Data in the Cloud can be in various states: at-rest, in-use, in-transit. The data do not have the same level of security requirements. Data being processed can not be protected with the same means as data in transit or at rest.

The primary method used to protect the transmitted and stored data is encryption. This method is still valid today for Cloud environment. Yet, this solution is not always possible as regard data-at-rest; In fact, a simple data encryption in an IaaS service is possible. However, data encrypting in a PaaS or SaaS application is not always possible. Data-at-rest used by Cloud applications is usually unencrypted because encryption prevents data indexing and searching. This is the case for data-in use that must be in a clear form for many applications [5, 7, 18, 25]. In 2009, IBM announced the development of complete homomorphic encryption that allows data to be processed by applications without their decrypting. The major inconvenient of this technique is his cost and computational complexity.

Concerning data-in-transit, the use of encryption alone is not sufficient to secure this type of data; Indeed, encryption guarantees confidentiality but not integrity [18]. Therefore, encryption algorithms are generally coupled with security protocols as well as network security equipment [5, 14, 18, 22]. Moreover, encryption techniques for data-at-rest and the data-in-transit may be different. For example, the encryption keys for the data in transit may be of short duration, while the keys for the data-at-rest may be preserved during a longer period [31].

Otherwise, data security solutions depend on other parameters such as: data size and data type. Indeed, the conventional solutions for securing a small set of data may be not suitable for large volume of data, like encryption and anonymization. Regarding data type, for example, sensitive data require confidentiality while personnel data require privacy; the information that must be protected in sensitive data is the content of data that were generally achieved by encryption techniques. Information that must be protected in personnel data is the user identity which generally achieved by anonymization techniques. All data type require more or less availability and integrity.

We summarized in Table 2 the common solutions used to secure data in Cloud Computing according to the parameters mentioned above. Our objective being to highlight the relationship between the different categories belonging to our classification and the influence of security on each other.

We can conclude from this table that Cloud characteristics have an impact on the data-at-rest, data-in-use and data-in-transit. This impact may affect their confidentiality, integrity as well as availability. The solutions mentioned in the table are not exhaustive; we focus only on the most known and widely used.

We have established a classification of common security techniques according to data security attributes, as shown in Fig. 2. The most important of them (The basic techniques) have been extracted and summarized from different literature works [2, 3, 6, 8, 12, 20].

These techniques, which are based for the most on simple encryption (symmetric and asymmetric), are often combined and adapted to form other models in order to meet the security requirements.

Trust and legal issues are another important concerns in the cloud.

Trust is one of the most difficult aspects to achieve. Leaving data to a third party is difficult to accept for users, especially, when data is sensitive [8, 15, 19]. Security measures appear insufficient in the absence of trust. Trust can be enhanced by security policies, provider transparency and introduction of a trusted third party in organization; however, even if the services are controlled and managed by authorized persons the physical infrastructure remains under the control of the provider. Generally, users trust their provider to ensure their data integrity and availability; however, they are more reticent about their data confidentiality and privacy.

**Table 2.** Main common solutions of data security according to Cloud characteristics, data life cycle and data security attributes

| Characteristics | Data states | | | Data security attributes | | | Main common solutions to data characteristics |
|---|---|---|---|---|---|---|---|
| | At-rest | In-use | In-transit | Confidentiality/privacy | Integrity | Availability | |
| Leased infrastructure | Impact | Impact | Impact | Impact | Impact | Impact | Encryption, access control, better transparency of service provider |
| Open infrastructure | Impact | Impact | Impact | Impact | Impact | Impact | Encryption, Access control |
| Shared infrastructure | Impact | Impact | Impact | Impact | Impact | Impact | Encryption, Access control |
| Elastic infrastructure | Impact | Impact | Impact | Impact | Impact | Impact | Encryption |
| Virtualization | Impact | Impact | Impact | Impact | Impact | Impact | VM securing, hypervisor securing, VPN |
| Distributed infrastructure | Impact | Impact | Impact | Impact | Impact | Impact | Better transparency of service provider, anonymization techniques |
| Main common solutions to data state/security attributes | Encryption, access control | Encryption, anonymization | Encryption, network security equipment, security protocols | Encryption, Access control, anonymization, data concealment | Integrity technique verification, Encryption | Resources/data duplication, backup | |

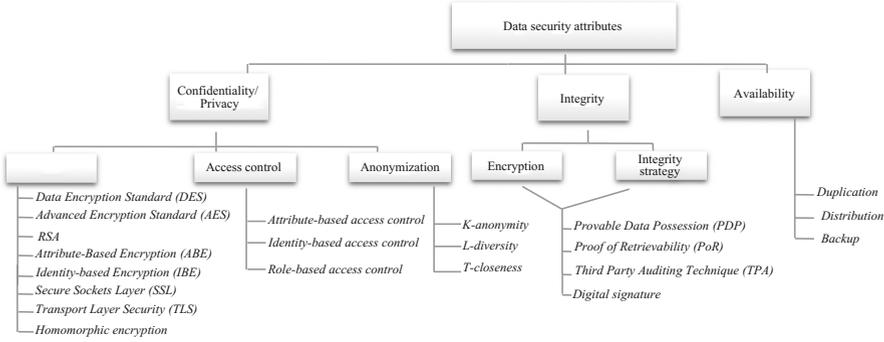

**Fig. 2.** Common data security techniques associated to data security attributes

Legal issues refer to confidentiality/privacy laws and regulations in the countries where the datacenters hosting the data are located [5, 8, 10, 14, 21, 24, 31]. Given that the Cloud is an emerging field, there is a lack of consensus on data security and privacy [3]. Privacy laws and regulations are obsolete and are no longer applicable to this new mode of data management and processing by a third party especially when data is sensitive [26, 32].

## 4 Conclusions

Cloud Computing presents numerous benefits compared to traditional infrastructure. Today, it is no longer important to understand what Cloud brings to the user, but rather to understand the issue related to the security of data hosted by a third party.

This paper discuss the issue of data security in Cloud Computing. We have, firstly, classify the data security issues according three dimensions: Cloud characteristics, data life cycle, and data security attributes. We have then identify the common solutions used to secure data for each category of this classification.

From our classification, it appears that data security issues depend on a number of criteria: data size (small/large), data type (personal/private/usage and identification) and data state (used, stocked or transferred). These criteria are influenced by single characteristics of the Cloud which depend in their turns of Cloud type/service used. These characteristics may affect the confidentiality, integrity and availability of such data.

We have also observed that the common solutions used to secure data in Cloud infrastructure such as encryption and access control are the classical solutions used to secure data in a traditional environment. These solutions are often combined and/or adapted to the cloud.

Ensuring data security in the Cloud cannot be summarized as a technical security solutions, it relies also on trust degree that user has in his provider as well as legal issues. Protection and management of intellectual property rights is an important issue: the transfer, storage and processing of data in Cloud Computing context require determination of solutions for the protection of intellectual property rights on these data. The simple question of who owns the data in the Cloud? Is problematic. According to European law, the data entrusted to a third party remain the property of the client and the law prohibits the provider from disclosing them. However, in Anglo-Saxon countries, the provider becomes the owner of the information.

Therefore, cloud security requires a deep questioning to current security measures. Once the data is on the Cloud, its management and control are transferred to the service providers. The idea that data is managed by a third party is not yet accepted, especially for large companies.